\documentclass[fleqn,10pt,a4paper]{article}
\usepackage{amsmath,amssymb,bm}
\usepackage{cite,epsfig,color,graphicx}

\title{Axionic suppression of plasma wakefield acceleration}

\author{D. A. Burton\thanks{Department of Physics, Lancaster University, Lancaster, UK and Cockcroft Institute of Accelerator Science and Technology, Daresbury, UK.}\and
A. Noble\thanks{Department of Physics, SUPA and University of Strathclyde, Glasgow, Scotland.}\and
T. J. Walton\footnotemark[1]\,\,\thanks{Current address: Department of Mathematics, University of Bolton, Bolton, UK.}}
\begin{document}
\maketitle
\begin{abstract}
Contemporary attempts to explain the existence of ultra-high energy cosmic rays using plasma-based wakefield acceleration deliberately avoid non-Standard Model particle physics. However, such proposals exploit some of the most extreme environments in the Universe and it is conceivable that hypothetical particles outside the Standard Model have significant implications for the effectiveness of the acceleration process. Axions solve the strong CP problem and provide one of the most important candidates for Cold Dark Matter, and their potential significance in the present context should not be overlooked. Our analysis of the field equations describing a plasma augmented with axions uncovers a dramatic axion-induced suppression of the energy gained by a test particle in the wakefield driven by a particle bunch, or an intense pulse of electromagnetic radiation, propagating at ultra-relativistic speeds within the strongest magnetic fields in the Universe.
\end{abstract}

\section{Introduction}
%%%%
The existence of cosmic rays with energies $\gtrsim 10^{20}\,{\rm eV}$ remains an outstanding puzzle in astrophysics~\cite{kotera:2011, letessier-selvon:2011}. The well-established Fermi paradigm for cosmic acceleration, in which charged particles gain energy by scattering repeatedly from magnetic inhomogeneities~\cite{fermi:1949} or shock waves~\cite{axford:1977, krymsky:1977, bell:1978, blandford:1978}, has not yet been shown to be efficient for accelerating particles to such exceptional energies. However, it has been argued that strategies whose effectiveness has been honed in the laboratory could be important in the astrophysical context~\cite{chen:2002}. Experience gained in the laboratory suggests that efficient acceleration mechanisms exploit sub-luminal waves, avoid collisional processes and do not involve particle trajectories with large curvature (thus avoiding excessive energy losses due to inelastic scattering and synchrotron radiation, respectively).  

A sufficiently short and intense laser pulse, or charged particle bunch, propagating rectilinearly through a laboratory plasma will drive an inhomogeneity in the plasma electron density that trails behind the pulse or bunch. The inhomogeneity, or wakefield, is a wave in the plasma electron density that propagates at the group velocity of the pulse (or velocity of the bunch) and whose longitudinal electric field is several orders of magnitude greater than that achievable within radio-frequency cavities. Driven mainly by the spectacular success of a number of landmark experiments and simulations~\cite{mangles:2004, geddes:2004, faure:2004, blumenfeld:2007} approximately a decade ago, plasma-based wakefield acceleration~\cite{tajima:1979} is now recognised as a vital concept for the next generation of terrestrial particle accelerators~\cite{assman:2014, xia:2014}.

In tandem, plasma wakefield acceleration has also gained attention in astrophysical contexts~\cite{chen:2002, diver:2010, ebisuzaki:2014}. For example, it has been argued that longitudinal space charge waves excited by Alfv\'en shocks caused by the collision of a pair of neutron stars may have the necessary properties for reliably accelerating protons to ${\rm ZeV}$ energies~\cite{chen:2002}. Thus, plasma wakefield acceleration may provide an effective solution to the problem of the origin of ultra-high energy cosmic rays.

In fact, neutron stars have long been suspects in the search for the culprits behind ultra-high energy cosmic rays~\cite{gunn:1969}.  In particular, the fast rotation of a pulsar combined with its strong magnetic field yields a particle accelerator based on unipolar induction, and the implications of this mechanism have been thoroughly explored over many years (see, e.g., reference~\cite{kotera:2011}). Although difficult questions remain over whether this approach can correctly explain the ultra-high energy cosmic ray spectrum~\cite{kotera:2011}, some success has been achieved in recent years in explaining the acceleration of heavy nuclei using sufficiently young strongly-magnetised neutron stars with millisecond rotation periods~\cite{fang:2012}. Nevertheless, plasma wakefield acceleration is an important alternative to unipolar induction, especially for light nuclei~\cite{chen:2002, ebisuzaki:2014}, and the focus of the present article will be on plasma wakefield acceleration in ultra-strong magnetic fields.

A strength of the plasma wakefield mechanism, like the unipolar inductor, for cosmic acceleration is that it does not rely on ingredients outside of the Standard Model of particle physics. However, non-Standard Model effects may be relevant as a consequence of the ultra-strong electromagnetic field strengths expected in the most extreme astrophysical environments. In particular, it is conceivable that effects due to hypothetical particles with very weak coupling to light and matter may manifest in such environments. One of the most popular Dark Matter candidates, the axion, was proposed as an elegant solution to the strong CP problem in QCD~\cite{peccei:1977, weinberg:1978, wilczek:1978} before its significance in the cosmological context was expounded~\cite{preskill:1983, abbott:1983, dine:1983}. Furthermore, in addition to the QCD axion, light pseudo-scalar particles are a generic consequence of type IIB string theory~\cite{cicoli:2012}; hence, a range of experiments have been developed, or are under development, in an attempt to uncover the effects of axions and axion-like particles (ALPs)~\cite{baker:2013, rosenberg:2015, mendonca:2007, dobrich:2010, villalba-chavez:2013, villalba-chavez:2014, seviour:2014}. Although positive detection remains elusive in the laboratory, it is possible that ALPs play a significant role in the ultra-strong magnetic field of a neutron star.

The purpose of this article is to demonstrate that ALPs may have significant consequences for plasma wakefield acceleration in ultra-strong magnetic fields. Although ALP fields couple directly to ordinary matter, the effects of the ALP-photon coupling are expected to dominate over those due to ALP-matter couplings in strongly magnetised plasmas. For convenience, we will briefly describe the main ingredients of our approach and summarise the key results before turning to their derivation.
\section{Ingredients in the model and key results}
%%%%
\label{sec:results}
Our assessment of the influence of ALPs is based on a model of a magnetised plasma that includes two charged pressureless perfect fluids. One fluid represents mobile electrons whilst the other fluid describes the charge carriers of a neutralising background medium with constant proper number density. The charge-to-mass ratio of the charge carriers of the background is assumed to be considerably lower than that of the other charged particles in the system and we assume that the motion of the background can be neglected over the timescales of interest.

The explanations in references~\cite{chen:2002, ebisuzaki:2014} of the origin of ultra-high energy cosmic rays invoke acceleration in astrophysical jets where the magnetic field is almost certainly too weak for ALPs to play a significant role. However, it has also been argued~\cite{diver:2010} that plasma-based acceleration is of interest within neutron stars, where the magnetic field is considerably higher.

The outer crust of a neutron star provides a background medium comprised of magnetically polarised iron atoms. Due to the considerable strength of the magnetic field ($\sim 10^8\,{\rm T}$), the electron `gas' within the crust is essentially confined to move along magnetic flux tubes threaded by the field lines~\cite{diver:2010}. Thus, we model the plasma wakefield as a sub-luminal non-linear longitudinal plane wave in the electron fluid density propagating at velocity $v$ parallel to the field lines of a homogeneous magnetic field of strength $B$ in the frame of the background medium. The electron density wave generates an ALP field whose strength is proportional to $g B$, where $g$ is the ALP-photon coupling constant; in turn, stress-energy-momentum conservation requires that the plasma fields are influenced by the ALP field. A self-consistent analysis of this type was previously used to uncover a novel signature of the ALP-photon coupling in waveguide mode spectra~\cite{noble:2008}.

For sufficiently large density fluctuations, the electric field of the wave has a sawtooth-like profile and its amplitude saturates; this leads to an upper bound on the energy that can be gained by a test particle in the wakefield. In particular, we will show that a test particle with mass $M$ and charge $Q$ cannot achieve an energy greater than $W^{\rm max} + Mc^2$ where
\begin{equation}
\label{energy_gain_ion_frame_dimensions_axion}
W^{\rm max} \approx W^{\rm max}_{B=0} \bigg[1 - \frac{\hbar^3}{\mu_0 c}\frac{g^2 B^2}{m_\alpha^2} \bigg(1-\frac{\tanh\sigma}{\sigma}\bigg)\bigg]
\end{equation}
with
\begin{equation}
\label{W_max}
W^{\rm max}_{B=0} = M c^2 \big[ 2\Theta^2(\gamma^3 - \gamma) + 2\Theta (\gamma^2 - 1)\sqrt{1 + \Theta^2(\gamma^2 - 1)} + \gamma - 1\big]
\end{equation}
the increase in energy in the absence of the magnetic field. The dimensionless parameters $\sigma$, $\Theta$ are
\begin{equation}
\label{s_expression_approx_l}
\sigma = \frac{\sqrt{2}\,m_\alpha c^2}{\hbar \omega_p} \gamma^{3/2},\qquad \Theta = \sqrt{\frac{|Q|\,m_{\rm e}}{e M}}
\end{equation}
where $m_\alpha$ is the ALP mass, $\gamma = 1/\sqrt{1-v^2/c^2}$ is the Lorentz factor of the phase speed $v$ of the wave and $\omega_p = \sqrt{e\, q_0 n_0/(\varepsilon_0 m_{\rm e})}$ is the plasma frequency ($q_0$ is the charge on a particle of the background medium, $n_0$ is the proper number density of the background medium, $e$ is the elementary charge and $m_{\rm e}$ is the mass of an electron). Equation (\ref{W_max}) yields the well-established result
\begin{equation}
W^{\rm max}_{B=0} = m_{\rm e} c^2 (4\gamma^3 - 3\gamma - 1)
\end{equation}
when the test particle is an electron ($M=m_{\rm e}$, $Q=-e$)~\cite{esarey:1995, esarey:2009}. 

The influence of the ALP field emerges as an overall multiplicative factor in (\ref{energy_gain_ion_frame_dimensions_axion}) that does not depend on the properties of the test particle. It is clear that $W^{\rm max} \lesssim W^{\rm max}_{B=0}$ where the approximate bound is saturated when $B = 0$, and the ALP field and the background magnetic field together reduce the maximum overall energy gain of the test particle. The previous result is reassuring because the energy supplied by the driver of the wakefield is shared between the ALP field and the plasma, and the test particle does not directly couple to the ALP field in our model. Furthermore, as we will show, QED vacuum polarisation does not contribute to (\ref{energy_gain_ion_frame_dimensions_axion}), to first order in small quantities, even though it affects the wavelength and amplitude of the wakefield.

It is important to note that our analysis does not incorporate the structure and dynamics of the driver of the wakefield. Neither the back-reaction of the plasma and ALP field on the driver, nor the back-reaction of the accelerated particle on the plasma, ALP field and driver, are included. Detailed investigation of such effects requires intensive numerical calculations that are beyond the scope of the present article, but it is likely that they would further reduce the energy gain of the accelerated bunch. For example, in most laboratory-based configurations, the coupling between the driver and the plasma in particle-beam driven wakefield acceleration imposes severe restrictions on the energy gain~\cite{chen:1986}.

The scale of the ALP-induced effects in (\ref{energy_gain_ion_frame_dimensions_axion}) can be estimated as follows.  The DESY ALPS-I (Any Light Particle) experiment has excluded light ALPs with strengths $g \gtrsim 10^{-7}\,{\rm GeV}^{-1}$, whilst astrophysical and cosmological considerations lead to $10^{-6}\,{\rm eV} \lesssim m_\alpha c^2 \lesssim 10^{-2}\,{\rm eV}$ for the QCD axion~\cite{villalba-chavez:2013}. The plasma frequency of the magnetically polarised iron lattice providing the neutralising background within the outer crust of a neutron star is $\sim 10^{18}\,{\rm Hz}$ when $B \sim 10^8\,{\rm T}$~\cite{diver:2010}. Hence, convenient estimates of the dimensionless parameters that characterise the ALP-induced suppression of a plasma wakefield accelerator driven by a $\sim 1\,{\rm TeV}$ bunch of electrons ($\gamma \sim 2 \times 10^6$) propagating along the $\sim 10^8\,{\rm T}$ magnetic field within the outer crust of a neutron star are as follows:
\begin{align}
\label{W_const}
&\frac{\hbar^3}{\mu_0 c}\frac{g^2 B^2}{m_\alpha^2} = 3.8\times 10^{-2}\,\bigg(\frac{g}{10^{-7}\,{\rm GeV^{-1}}}\bigg)^2\bigg(\frac{B}{10^8\,{\rm T}}\bigg)^2\bigg(\frac{10^{-5}\,{\rm eV}}{m_\alpha c^2}\bigg)^2,\\
\label{sigma_val}
&\sigma =  9.7\,\bigg(\frac{m_\alpha c^2}{10^{-5}\,{\rm eV}}\bigg)\bigg(\frac{2\pi \times 10^{18}\,{\rm rad\,s^{-1}}}{\omega_p}\bigg) \bigg(\frac{\gamma}{2\times 10^6}\bigg)^{3/2}.
\end{align}

Note that a $\sim 3\%$ perturbation to $W^{\rm max}_{B=0}$ follows from (\ref{energy_gain_ion_frame_dimensions_axion}), (\ref{W_const}), (\ref{sigma_val}) when the representative parameters are used. The relative size of the perturbation is highly sensitive to the ALP mass; in particular, it reduces to $\sim 0.04\%$ for $m_\alpha c^2 = 10^{-4}\,{\rm eV}$ and increases to $\sim 90\%$ for $m_\alpha c^2 = 10^{-6}\,{\rm eV}$. As we will see in the remainder of this article, a perturbative approach is used to obtain (\ref{energy_gain_ion_frame_dimensions_axion}) and, although the use of perturbation theory is suspect when $m_\alpha c^2 \lesssim 10^{-6}\,{\rm eV}$, it is reasonable to conclude that the ALP field has a substantial effect on the maximum energy gain if $m_\alpha c^2 < 10^{-5}\,{\rm eV}$.

More stringent upper bounds on the ALP-photon coupling constant $g$, such as those obtained from solar axion searches by CAST~\cite{barth:2013} ($g \lesssim 10^{-10}\,{\rm GeV}^{-1}$), can be compensated by the magnetic fields found in magnetars ($B \sim 10^{11}\,{\rm T}$). The plasma frequency of the outer crust of a neutron star satisfies $\omega_p\sim 2\pi \times 10^{18}\,(B/ 10^8\,{\rm T})^\delta\,{\rm rad\,s^{-1}}$, where $\delta \sim 3/5 - 3/4$~\cite{diver:2010}, and inspection of (\ref{sigma_val}) shows that a wakefield driven by a $\sim 30\,{\rm TeV}$ bunch of electrons in a $\sim 10^{11}\,{\rm T}$ field leads to values of $\sigma$ that are similar to those found when $B \sim 10^8\,{\rm T}$. Note that only relatively modest drive-bunch energies (in the context of cosmic rays) are required to obtain substantial axionic suppression of the plasma wakefield acceleration mechanism. 
\section{Derivation of the results}
%%%%
Henceforth, units are used in which the speed of light $c$, the permittivity $\varepsilon_0$ of free space and the reduced Planck constant $\hbar$ are unity. Furthermore, for linguistic convenience, we will refer to the charge carriers of the neutralising background medium as {\it ions} regardless of whether they are polarised ion cores in the crust of a neutron star or ions in a magnetised plasma. 
%%%
\subsection{Field equations}
%%%%
The variables describing a cold {\it ALP-plasma} are the ALP $0$-form $\alpha$, the electromagnetic $2$-form $F$, the $4$-velocity field $V_{\rm e}$ of the plasma electrons, the proper number density $n_{\rm e}$ of the plasma electrons, the $4$-velocity field $V_{\rm 0}$ of the ions and the proper number density $n_0$ of the ions. The effect of the electromagnetic field on the ions is negligible over the length and time scales of interest; we choose $n_0$ to be constant and choose $V_{\rm 0}=\partial/\partial t$ where the spacetime metric $\eta$ and volume $4$-form $\star 1$ are
\begin{align}
&\eta = - dt \otimes dt + dx \otimes dx + dy \otimes dy + dz \otimes dz,\\
&\star 1 = dt\wedge dx\wedge dy\wedge dz
\end{align}
with $\wedge$ the exterior product and $d$ the exterior derivative on differential forms. Exterior differential calculus is used extensively in this section because it is an efficient tool for formulating the ALP-plasma field equations and reducing them to non-linear ODEs; a detailed account of the techniques and conventions used here may be found in reference~\cite{burton:2003}.

The Hodge map $\star$ induced from the volume $4$-form $\star 1$ satisfies the identity $\star(\beta \wedge \widetilde{W}) = \iota_W \star\beta$ for all differential forms $\beta$ and vectors $W$. The action of $\star$ is extended to non-decomposable forms by linearity, and the linear operator $\iota_W$ is the interior derivative with respect to $W$. The $1$-form $\widetilde{W}$ is the metric dual of $W$ and satisfies $\widetilde{W}(V) = \eta(W,V)$ for all vectors $V$; likewise, the vector field $\widetilde{\beta}$ is the metric dual of the $1$-form $\beta$ and satisfies $\widetilde{V}(\widetilde{\beta}) = \beta(V)$ for all vectors $V$.

The electric $4$-current densities of the plasma electrons and the ion background are $q_{\rm e} n_{\rm e} V_{\rm e}$, $q_0 n_0 V_0$, respectively, where $q_{\rm e} = -e$ is the charge on an electron, $q_0$ is the charge on a background ion and $\eta(V_{\rm e},V_{\rm e}) = \eta(V_{\rm 0},V_{\rm 0}) = -1$ . The behaviour of the electromagnetic field is determined by the Gauss-Faraday and Gauss-Amp\`ere laws:
\begin{align}
\label{maxwell_F}
&dF = 0,\\
\label{maxwell_G}
&d\star G = - q_{\rm e} n_{\rm e}\star\widetilde{V_{\rm e}} - q_0 n_0\star\widetilde{V_{\rm 0}}
\end{align}
and the ALP field $\alpha$ satisfies
\begin{equation}
\label{axion_field_eqn}
d\star d\alpha - m^2_{\alpha} \alpha \star 1 = - \partial_\alpha \lambda \star 1
\end{equation}
with $m_\alpha$ the ALP mass. The electromagnetic excitation $2$-form $G$ is specified by
\begin{equation}
\label{excitation_2-form}
G = 2\big(\partial_X\lambda\,F - \partial_Y\lambda\,\star F\big)
\end{equation}
where $\lambda$ is a $0$-form-valued function of the ALP field $\alpha$ and the electromagnetic invariants
\begin{equation}
X = \star(F\wedge\star F),\quad Y=\star(F\wedge F).
\end{equation}
The $0$-form $\lambda(X,Y,\alpha)$ is the sum of all purely electromagnetic contributions to the Lagrangian (including effective self-couplings due to QED vacuum polarisation) and terms that encode the interaction between the electromagnetic and ALP fields.   

The behaviour of the electron fluid is determined by appealing to total stress-energy-momentum conservation. The divergence of the total stress-energy-momentum tensor of the electron fluid, electromagnetic field and ALP field must balance the forces on the ions. For present purposes, it is useful to cast this statement in the language of differential forms; the behaviour of the plasma electrons is determined by the following field equation:
\begin{equation}
\label{stress_balance}
d\tau_K = q_0 n_0 \iota_K F\wedge\star\widetilde{V_0}
\end{equation}
where the total stress $3$-form $\tau_K$ of the electromagnetic field, ALP field and plasma electron fluid is
\begin{align}
\label{stress_3-form}
\tau_K = \iota_K F \wedge \star G + \lambda \star \widetilde{K} + \frac{1}{2}\big(\iota_K d\alpha\wedge\star d\alpha + d\alpha\wedge\iota_K\star d\alpha - m^2_{\rm \alpha} \alpha^2 \star \widetilde{K} \big)
+ m_{\rm e} n_{\rm e}\, \eta(V_{\rm e},K) \star\widetilde{V_{\rm e}}
\end{align}
with $K$ a Killing vector and $m_{\rm e}$ the mass of an electron. The stress $3$-form $\tau_K$ is related to the total stress-energy-momentum tensor ${\cal T}$ of the electromagnetic field, ALP field and plasma electron fluid via the identity ${\cal T}(K,W) = \iota_W \star\tau_K$, for all vector fields $W$, and (\ref{stress_balance}) is the component of a local balance law associated with $K$. In particular, if $K$ is a generator of spatial translations then (\ref{stress_balance}) is the $K$-component of a field equation describing the local balance of linear momentum whilst a generator of time translations leads to local energy balance.

Before reducing the above field equations to a system of non-linear ODEs, it is worth commenting on the approach adopted here in comparison to our earlier analysis of plasma wakefields in non-linear electrodynamics~\cite{burton:2011a}. It can be shown that the system  (\ref{maxwell_F}), (\ref{maxwell_G}), (\ref{axion_field_eqn}), (\ref{stress_balance}) is equivalent to (\ref{maxwell_F}), (\ref{maxwell_G}), (\ref{axion_field_eqn}) with the Lorentz equation of motion for the plasma electron fluid:
\begin{equation}
\label{lorentz_force}
\iota_{V_{\rm e}}d\widetilde{V_{\rm e}} = \frac{q_{\rm e}}{m_{\rm e}} \iota_{V_{\rm e}} F.
\end{equation}
The system (\ref{maxwell_F}), (\ref{maxwell_G}), (\ref{lorentz_force}) is the starting point for the analysis in reference~\cite{burton:2011a}; however, the analysis in reference~\cite{burton:2011a} includes the derivation of a first integral that can be shown to follow immediately from the stress-energy-momentum balance law (\ref{stress_balance})~\cite{burton:2015}. For present purposes, it is more efficient to adopt the strategy that we recently developed in reference~\cite{burton:2015} and begin with the stress-energy-momentum balance law (\ref{stress_balance}) rather than (\ref{lorentz_force}).
\subsection{ODE system describing non-linear longitudinal ALP-plasma waves}
%%%%
Solutions to (\ref{maxwell_F}), (\ref{maxwell_G}), (\ref{axion_field_eqn}), (\ref{stress_balance}) are sought that describe longitudinal plane waves propagating along a constant magnetic field. The electromagnetic field $F$ and electron fluid $4$-velocity $V_{\rm e}$ have the form
\begin{align}
\label{F_ansatz}
&F = E(\zeta)\,dt\wedge dz - B\,dx\wedge dy,\\
\label{V_ansatz}
&\widetilde{V_{\rm e}} = \mu(\zeta)\,\theta^1 - \sqrt{\mu(\zeta)^2 - \gamma^2}\,\theta^2
\end{align}
where $\zeta = z - v t$ with $v$ the phase speed of the wave, $0 < v < 1$, $\gamma=1/\sqrt{1-v^2}$, $\theta^1 = - dt + v dz$  and $\theta^2 = d\zeta = dz - v dt$. Likewise, the electron proper number density $n_{\rm e}$ and the ALP field $\alpha$ are assumed to depend on $\zeta$ only.

Clearly, the Gauss-Faraday law (\ref{maxwell_F}) is trivially satisfied by (\ref{F_ansatz}). Furthermore, the point-wise behaviour of the components of $\star\,G$ with respect to the co-frame $\{dt, dx, dy, dz\}$ depends on $\zeta$ only and it immediately follows that $d\zeta \wedge d\star G = 0$. Thus, the exterior product of $d\zeta$ and the Gauss-Amp\`ere law (\ref{maxwell_G}) yields
\begin{equation}
\label{electron_density}
n_{\rm e} = -\frac{q_0}{q_{\rm e}} n_0 v\gamma^2 \frac{1}{\sqrt{\mu^2-\gamma^2}}.
\end{equation}

The set of Killing vectors $\{\partial/\partial x, \partial/\partial y, \widetilde{\theta^1}, \widetilde{\theta^2}\}$ is a basis for vector fields on flat spacetime. Setting $K$ to $\partial/\partial x$ or $\partial/\partial y$ in (\ref{stress_3-form}) yields $\tau_K \simeq 0$ where $\simeq$ denotes equality modulo exact forms. On the other hand, inspection of (\ref{F_ansatz}) reveals that the right-hand side of (\ref{stress_balance}) vanishes when $K\in\{\partial/\partial x, \partial/\partial y\}$ and so (\ref{stress_balance}) is trivially satisfied.

Setting $K=\widetilde{\theta^1}$ in (\ref{stress_3-form}) reveals
\begin{equation}
\label{tau_theta1_simplified}
\tau_K \simeq - m_{\rm e}\frac{q_0}{q_{\rm e}}n_0 v \mu\, \theta^1\wedge dx \wedge dy
\end{equation}
where (\ref{electron_density}) has been used to eliminate $n_{\rm e}$, and
\begin{equation}
\label{E_mu}
E = \frac{m_{\rm e}}{q_{\rm e}} \frac{1}{\gamma^2} \mu^\prime
\end{equation}
follows from (\ref{stress_balance}), (\ref{F_ansatz}), (\ref{tau_theta1_simplified}). A prime indicates differentiation with respect to $\zeta$.

The remaining outcome of (\ref{stress_balance}) is revealed by choosing $K=\widetilde{\theta^2}$. It can be shown that
\begin{align}
\label{tauK_e2}
\tau_K \simeq &\bigg[2(-\partial_X\lambda\,E^2 - \partial_Y\lambda\,EB)+\lambda+\frac{1}{2}\bigg(\frac{\alpha^{\prime\,2}}{\gamma^2} - m^2_\alpha\alpha^2\bigg) 
- \frac{q_0}{q_{\rm e}}m_{\rm e}n_0 v\sqrt{\mu^2-\gamma^2}\bigg] \theta^1\wedge dx \wedge dy
\end{align}
where (\ref{electron_density}) has been used to eliminate $n_{\rm e}$. Thus, combining (\ref{tauK_e2}) with the result of using (\ref{E_mu}) to eliminate $E$ in the right-hand side of (\ref{stress_balance}) reveals
\begin{align}
\label{tauK_e2_ODE}
0 = \bigg[2(\partial_X\lambda\,E^2 + \partial_Y\lambda\,EB)-\lambda-\frac{1}{2}\bigg(\frac{\alpha^{\prime\,2}}{\gamma^2} - m^2_\alpha\alpha^2\bigg)
 + \frac{q_0}{q_{\rm e}}m_{\rm e}n_0 (v\sqrt{\mu^2-\gamma^2}-\mu)\bigg]^\prime.
\end{align}
Only the ALP field equation (\ref{axion_field_eqn}) remains and it can be shown that
\begin{equation}
\label{axion_ODE}
\frac{\alpha^{\prime\prime}}{\gamma^2} - m^2_\alpha \alpha = - \partial_\alpha \lambda.
\end{equation}
It is worth noting that the above procedure bypasses the second-order ODE for $\mu$ that emanates from the Gauss-Amp\`ere law (\ref{maxwell_G}). However, the latter ODE also follows as a consequence of (\ref{E_mu}), (\ref{tauK_e2_ODE}), (\ref{axion_ODE}) and no new information is uncovered. 

In summary, the ALP-plasma field system reduces to (\ref{tauK_e2_ODE}), (\ref{axion_ODE}) with $E$ given by (\ref{E_mu}). The electromagnetic invariants are $X=E^2-B^2$, $Y=2EB$. For definiteness, from now on we will only consider theories with a minimal coupling between the ALP and the electromagnetic field rather than those with more general couplings~\cite{villalba-chavez:2013, burton:2011b}. We will focus on theories of the form
\begin{equation}
\label{lambda_axion}
\lambda(X,Y,\alpha) = \lambda_{\rm EM}(X,Y) + \frac{1}{2}g\alpha Y
\end{equation}
where $\lambda_{\rm EM}(X,Y)$ depends only on the electromagnetic field invariants and $g$ is a coupling constant. The term $g\alpha Y/2$ coupling the ALP field and electromagnetic field does not appear in the stress-energy-momentum tensor and its contributions to the first three terms in (\ref{tauK_e2_ODE}) cancel; substituting $\lambda$ in (\ref{tauK_e2_ODE}), (\ref{axion_ODE}) leads to
\begin{align}
\label{tauK_e2_ODE_axion}
0 = \bigg[2(\partial_X\lambda_{\rm EM}\,E^2 + \partial_Y\lambda_{\rm EM}\,EB)-\lambda_{\rm EM}-\frac{1}{2}\bigg(\frac{\alpha^{\prime\,2}}{\gamma^2} - m^2_\alpha\alpha^2\bigg) + \frac{q_0}{q_{\rm e}}m_{\rm e}n_0 (v\sqrt{\mu^2-\gamma^2}-\mu)\bigg]^\prime
\end{align}
and
\begin{equation}
\label{axion_ODE_axion}
\frac{\alpha^{\prime\prime}}{\gamma^2} - m^2_\alpha \alpha = - g E B.
\end{equation}
The $0$-form $\lambda_{\rm EM}(X,Y)$ reduces to the classical vacuum Maxwell Lagrangian $X/2$ in the weak-field limit and, for stronger fields, captures the effects of quantum vacuum polarisation. As we will soon see, the detailed structure of $\lambda_{\rm EM}(X,Y)$ is unimportant for our purposes.  
\subsection{Periodic solutions}
%%%%
Inspection of (\ref{axion_ODE_axion}) shows that, in general, $\alpha$ behaves exponentially when $g=0$ and such solutions describe ALP fields generated by a source other than the plasma electrons and ions. Although such effects could be attributed to the driver of the plasma wakefield, we will only consider the ALP fields self-consistently generated by the plasma electrons and ions.

The structure of (\ref{tauK_e2_ODE_axion}) suggests the existence of solutions for $\mu$ that are periodic in $\zeta$ when $\alpha=0$. Furthermore, the structure of (\ref{axion_ODE_axion}) ensures that an ALP field generated by a periodic electric field will be periodic; hence, in the absence of a background ALP field, a Fourier series representation for $\mu$, $\alpha$ will be sought:
\begin{align}
\label{mu_fourier}
&\mu(\zeta) = \sum\limits^\infty_{n=-\infty} \mu_n \exp\bigg(2\pi i n \frac{\zeta}{l}\bigg) ,\\
\label{alpha_fourier}
&\alpha(\zeta) = \sum\limits^\infty_{n=-\infty} \alpha_n \exp\bigg(2\pi i n \frac{\zeta}{l}\bigg) 
\end{align}
where the period $l$ of the solution must be determined as part of the analysis. Thus (\ref{E_mu}), (\ref{axion_ODE_axion}), (\ref{mu_fourier}), (\ref{alpha_fourier}) lead to the relationship
\begin{equation}
\label{alpha_n}
\alpha_n = \frac{gB l}{4\pi^2 n^2 + m^2_\alpha \gamma^2 l^2}\frac{m_{\rm e}}{q_{\rm e}} 2\pi i n \mu_n
\end{equation}
between the Fourier components of $\alpha$ and $\mu$, and we can express (\ref{tauK_e2_ODE_axion}) as the non-linear and non-local equation
\begin{align}
\label{tauK_e2_ODE_axion_eliminated}
0 = \bigg[2(\partial_X\lambda_{\rm EM}\,E^2 + \partial_Y\lambda_{\rm EM}\,EB)-\lambda_{\rm EM} + \varphi[\mu]
+ \frac{q_0}{q_{\rm e}}m_{\rm e}n_0 (v\sqrt{\mu^2-\gamma^2}-\mu)\bigg]^\prime
\end{align}
for $\mu$ where the functional $\varphi$ is
\begin{equation}
\label{axion_remnant_functional}
\varphi[\mu] = -\frac{1}{2}\bigg(\frac{\alpha^{\prime\,2}}{\gamma^2} - m^2_\alpha\alpha^2\bigg)
\end{equation}
with $\alpha$ specified by (\ref{alpha_fourier}), (\ref{alpha_n}) and
\begin{equation}
\mu_n = \frac{1}{l} \int^l_0 \exp(-2\pi i n \zeta/l)\,\mu(\zeta)\,d\zeta.
\end{equation}
\subsection{Maximum energy gain}
%%%%
Following the approach used in reference~\cite{burton:2015}, the change in energy $\Delta{\cal E}_K$ of a test particle between the endpoints ${\rm I}$, ${\rm II}$ of a segment $C$ of its world-line in an inertial frame of reference adapted to a timelike unit Killing vector $K$ is
\begin{align}
\nonumber
\Delta{\cal E}_K &= - M \eta(\dot{C},K)\big|_{\rm I}^{\rm II}\\
\label{covariant_work_done}
&= Q \int_C \iota_K F
\end{align}
where $M$, $Q$ are the mass and charge of the test particle, respectively, and $\dot{C}$ is the $4$-velocity of the particle. Equation (\ref{covariant_work_done}) is a covariant expression of the relationship between the change in energy of the particle and the work done on the particle by the Lorentz force between ${\rm I}$ and ${\rm II}$. Of particular interest here is the case when $\zeta_{\rm I}$, $\zeta_{\rm II}$ are located at adjacent turning points of $\mu$, i.e. adjacent nodes of the electric field.

The square root in (\ref{tauK_e2_ODE_axion_eliminated}) ensures that $\mu > \gamma$ and leads to an upper bound on the amplitude of the solution to (\ref{tauK_e2_ODE_axion_eliminated}) if $\lambda_{\rm EM} = X/2$ and $g=0$ (i.e. the effects of quantum vacuum polarisation and the ALP are neglected). However, the effects of quantum vacuum polarisation and the ALP are expected to be small and, from a perturbative perspective, an upper bound on $\mu$ is also expected to exist when such effects are included.

The electric field $\iota_K F$ in a frame adapted to $K = \gamma(\partial_t + v\partial_z)$ (i.e. a frame in which the wave is static) is $E\gamma\, d\zeta$ using (\ref{F_ansatz}) and $d\zeta=dz - vdt$. Thus, the maximum gain in energy of the test particle is
\begin{align}
\nonumber
\Delta{\cal E}_K &= \frac{Q}{\gamma}\frac{m_{\rm e}}{q_{\rm e}} \int^{\zeta_{\rm II}}_{\zeta_{\rm I}} \frac{d\mu}{d\zeta}\,d\zeta\\
\label{DEK_value}
&= \frac{Q}{\gamma}\frac{m_{\rm e}}{q_{\rm e}} (\mu_{\rm II} - \mu_{\rm I})
\end{align}
where (\ref{E_mu}) has been used. Although (\ref{DEK_value}) can be expressed in terms of the potential difference between two adjacent nodes of the electric field, it is convenient for present purposes to retain $\Delta{\cal E}_K$ in the form given above.

The parameters $\mu_{\rm I}$, $\mu_{\rm II}$ are adjacent turning points in $\mu$ and further analysis is required to determine them. Integrating (\ref{tauK_e2_ODE_axion_eliminated}) immediately yields
\begin{equation}
\label{integrated_tauK_e2_ODE_axion_eliminated}
\bigg\{\varphi[\mu] + \frac{q_0}{q_{\rm e}}m_{\rm e}n_0 (v\sqrt{\mu^2-\gamma^2}-\mu)\bigg\}\bigg|^{\zeta_{\rm II}}_{\zeta_{\rm I}} = 0
\end{equation}
because $E \propto \mu^\prime$ and so $E(\zeta_{\rm{I}}) = E(\zeta_{\rm{II}}) = 0$.  In order to ensure that the test particle is accelerated, rather than decelerated, we require the particle to be in the appropriate phase of the wave. If $Q/q_{\rm e} > 0$ then we choose $\mu_{\rm I}$ to be a minimum of $\mu$ and $\mu_{\rm II}$ to be the subsequent maximum of $\mu$. However, if $Q/q_{\rm e} < 0$ then we choose $\mu_{\rm I}$ to be a maximum of $\mu$ and $\mu_{\rm II}$ to be the subsequent minimum of $\mu$.

Note that non-linear corrections to $\lambda_{\rm EM} = X/2$ only contribute to (\ref{integrated_tauK_e2_ODE_axion_eliminated}) through $\varphi[\mu]$. However, as noted above, the corrections to classical Maxwell theory due to quantum vacuum polarisation and the coupling to the ALP are expected to be amenable to perturbative analysis. Since $\varphi[\mu]$ is a small perturbation to the remaining terms in (\ref{integrated_tauK_e2_ODE_axion_eliminated}), we expect the value of $\mu$ for a classical cold plasma to be adequate for calculating the first term in (\ref{integrated_tauK_e2_ODE_axion_eliminated}). Furthermore, inspection of (\ref{integrated_tauK_e2_ODE_axion_eliminated}) reveals that the absolute minimum value of $\mu$ is $\gamma$ when $\varphi[\mu]$ is neglected and this conclusion will hold when $\varphi[\mu]$ is not neglected. Thus, we express (\ref{integrated_tauK_e2_ODE_axion_eliminated}) as
\begin{equation}
\label{integrated_tauK_e2_ODE_axion_eliminated_approx}
\varphi[\nu]\big|^{\zeta_{\rm II}}_{\zeta_{\rm I}}  + \epsilon \frac{q_0}{q_{\rm e}}m_{\rm e}n_0 (v\sqrt{\mu_*^2-\gamma^2}-\mu_* + \gamma) \approx 0
\end{equation}
with $\nu$ the solution to (\ref{tauK_e2_ODE_axion}) for $\mu$ that arises when $\lambda_{\rm EM} = X/2$, $\alpha=0$. The parameters $\epsilon$, $\mu_*$ satisfy
\begin{equation}
\epsilon =
\begin{cases}
+1\,\,&\mbox{if $Q/q_{\rm e} > 0$}\\
-1\,\,&\mbox{if $Q/q_{\rm e} < 0$}
\end{cases}
\end{equation}
and
\begin{equation}
\mu_* =
\begin{cases}
\mu_{\rm II}\,\,&\mbox{if $Q/q_{\rm e} > 0$}\\
\mu_{\rm I}\,\,&\mbox{if $Q/q_{\rm e} < 0$}
\end{cases}
\end{equation}
with $\mu_{\rm I} = \gamma$ if $Q/q_{\rm e} > 0$ or $\mu_{\rm II} = \gamma$ if $Q/q_{\rm e} < 0$. Setting $\lambda_{\rm EM} = X/2$, $\alpha=0$ in (\ref{tauK_e2_ODE_axion}) reveals
\begin{equation}
\bigg[\frac{1}{2}\mu^{\prime 2}  - \gamma^4\omega^2_p (v\sqrt{\mu^2-\gamma^2}-\mu)\bigg]^\prime = 0
\end{equation}
where $\omega_p = \sqrt{-q_0 q_{\rm e} n_0/m_{\rm e}}$ is the plasma (angular) frequency, and it follows that $\nu$ satisfies
\begin{equation}
\frac{1}{2}\nu^{\prime 2}  - \gamma^4\omega^2_p (v\sqrt{\nu^2-\gamma^2}-\nu + \gamma) = 0
\end{equation}
since, by construction, $\nu^\prime = 0$ when $\nu=\gamma$.

Note that in calculating the {\it maximum} gain in energy, all electromagnetic self-couplings (such as those that arise from QED) are distilled away without further approximation. The explicit modifications to the period and profile of the wave due to vacuum polarisation conspire to produce no change in the maximum possible energy gain~\cite{burton:2015}. The effects of quantum vacuum polarisation implicit in the remnants of the ALP field in (\ref{integrated_tauK_e2_ODE_axion_eliminated}) are lost when $\varphi[\mu]$ is approximated by $\varphi[\nu]$.

Solving (\ref{integrated_tauK_e2_ODE_axion_eliminated_approx}) for $\mu_*$ and casting the result in an amenable form is a little involved and, to avoid distraction, we have presented the details in the Appendix. We find
\begin{equation}
\label{mu_II_explicit_g0}
\mu_*\big|_{g=0} = \gamma^3(1+v^2)
\end{equation}
in the absence of the ALP-photon coupling and
\begin{equation}
\label{mu_II_explicit}
\mu_* \approx \gamma^3(1+v^2) - \frac{2 g^2 B^2}{m_\alpha^2}\gamma^3 \bigg[1-\frac{2}{\pi s}\tanh\bigg(\frac{\pi s}{2}\bigg)\bigg]
\end{equation}
to first order in $g^2$, where the parameter $s$ is
\begin{equation}
\label{s_def}
s = \frac{m_\alpha \gamma l}{2\pi}
\end{equation}
and only the leading order $\gamma$-dependence of the ${\cal O}(g^2)$ term in (\ref{mu_II_explicit}) has been retained. Hence,  using (\ref{mu_II_explicit_g0}) and (\ref{DEK_value}) with $\mu_{\rm I} = \gamma$ if $Q/q_{\rm e} > 0$ or $\mu_{\rm II} = \gamma$ if $Q/q_{\rm e} < 0$, the maximum gain in energy in the wave frame when $g=0$ is
\begin{equation}
\label{energy_gain_wave_frame_g0}
\Delta{\cal E}_K\big|_{g=0} = 2 m_{\rm e}\gamma^2 v^2 \bigg|\frac{Q}{q_{\rm e}}\bigg|
\end{equation}
and
\begin{equation}
\label{energy_gain_wave_frame}
\Delta{\cal E}_K \approx 2 m_{\rm e}\gamma^2 \bigg|\frac{Q}{q_{\rm e}}\bigg|\bigg\{v^2 - \frac{g^2 B^2}{m_\alpha^2}\bigg[1-\frac{2}{\pi}\frac{1}{s}\tanh\bigg(\frac{\pi s}{2}\bigg)\bigg]\bigg\}
\end{equation}
follows from a perturbative analysis of (\ref{mu_II_explicit}), (\ref{DEK_value}) in $g^2$.

The maximum energy gain in the inertial frame in which the ions are at rest is straightforward to obtain by adapting the approach in reference~\cite{burton:2015} to accommodate a test particle with arbitrary mass $M$ and charge $Q$. Unlike (\ref{energy_gain_wave_frame}), the energy gain in the ion frame depends on the initial velocity of the test particle. Although it is intuitively obvious that the test particle should begin at rest in the wave frame to gain maximum energy within the wave, we will prove this result for completeness.

Expressing the $4$-velocity $\dot{C}$ of the test particle as $\dot{C} = \Gamma (K + U L)$ where $(0,0,U)$ is the $3$-velocity of the test particle in the wave frame, $K=\gamma(\partial_t + v \partial_z)$, $L=\gamma(\partial_z + v \partial_t)$ and $\Gamma = 1/\sqrt{1-U^2}$ gives
\begin{align}
\nonumber
\Delta{\cal E}_{\partial_t} &= -M\eta(\dot{C},\partial_t)|^{\rm II}_{\rm I}\\
\label{DeltaE_PDt}
&= M\gamma[\Gamma_{\rm II}(1+U_{\rm II} v) - \Gamma_{\rm I}(1+U_{\rm I}v)]
\end{align}
where $U_{\rm I}$, $U_{\rm II}$ are the values of $U$ and $\Gamma_{\rm I}$, $\Gamma_{\rm II}$ are the values of $\Gamma$ at the spacetime events ${\rm I}$, ${\rm II}$ respectively. The expression
\begin{align}
\nonumber
\Delta{\cal E}_K &= -M \eta(\dot{C},K)|^{\rm II}_{\rm I}\\
\label{DEK_gamma_II}
&= M (\Gamma_{\rm II} - \Gamma_{\rm I})
\end{align}
fixes $\Gamma_{\rm II}$ (and therefore also fixes $U_{\rm II}$) in terms of $\Gamma_{\rm I}$ and the known quantity $\Delta{\cal E}_K$. Since $\Delta {\cal E}_K$ is independent of $\Gamma_{\rm I}$ it follows that $d\Gamma_{\rm II}/d\Gamma_{\rm I} = 1$ using (\ref{DEK_gamma_II}) and
\begin{equation}
\frac{d\Delta{\cal E}_{\partial_t}}{d\Gamma_{\rm I}} = M \gamma v\bigg(\frac{1}{U_{\rm II}} - \frac{1}{U_{\rm I}}\bigg)
\end{equation}
follows from (\ref{DeltaE_PDt}); therefore $d\Delta{\cal E}_{\partial_t}/d\Gamma_{\rm I} < 0$ since $U_{\rm II} > U_{\rm I}$ (the test particle is assumed to be in the phase of the plasma wave in which it is accelerated). Hence, the maximum value $\Delta{\cal E}^{\rm max}_{\partial_t}$ of $\Delta{\cal E}_{\partial_t}$ is obtained at the minimum value of $\Gamma_{\rm I}$ and setting $\Gamma_{\rm I} = 1$ in (\ref{DeltaE_PDt}) yields
\begin{equation}
\label{Delta_E_max}
\Delta{\cal E}^{\rm max}_{\partial_t} = M\gamma[\Gamma_{\rm II}(1+U_{\rm II} v) - 1]
\end{equation}
where $\Gamma_{\rm II}$ is specified in terms of $\Delta{\cal E}_K$ as
\begin{equation}
\label{gamma_II}
\Gamma_{\rm II} = 1 + \frac{1}{M}\Delta{\cal E}_K.
\end{equation}

The maximum total energy that the test particle can achieve in the ion frame is $W^{\rm max} + M$ where
\begin{equation}
\label{energy_gain_ion_frame_exact}
W^{\rm max} = \Delta{\cal E}^{\rm max}_{\partial_t} + M (\gamma - 1).
\end{equation}
Using (\ref{energy_gain_wave_frame_g0}), (\ref{Delta_E_max}), (\ref{gamma_II}) to eliminate $\Delta{\cal E}^{\rm max}_{\partial_t}$ from (\ref{energy_gain_ion_frame_exact}) in favour of $\gamma$ and the ALP-plasma parameters leads to
\begin{equation}
\label{energy_gain_ion_frame_pre_final_s_g0}
W^{\rm max}\big|_{g=0} = M \big[ 2\Theta^2 (\gamma^3 - \gamma) + 2\Theta(\gamma^2 - 1)\sqrt{1+\Theta^2(\gamma^2-1)} + \gamma - 1\big]
\end{equation}
where
\begin{equation}
\label{theta_end}
\Theta = \sqrt{\bigg|\frac{Q}{q_{\rm e}}\bigg| \frac{m_{\rm e}}{M}}
\end{equation}
and a perturbative analysis of (\ref{Delta_E_max}) in $g^2$ yields
\begin{equation}
\label{energy_gain_ion_frame_pre_final_s_no_c_pre}
W^{\rm max} \approx W^{\rm max}\big|_{g=0} \bigg\{1 - \frac{g^2 B^2}{m_\alpha^2}\bigg[1-\frac{2}{\pi}\frac{1}{s}\tanh\bigg(\frac{\pi s}{2}\bigg)\bigg]\bigg\}
\end{equation}
for $\gamma \gg 1$. The equality in (\ref{energy_gain_ion_frame_pre_final_s_no_c_pre}) is approximate because it is only valid to first order in $g^2$ and the dominant behaviour of the ${\cal O}(g^2)$ term in large $\gamma$ has been exploited to write the final result in a factored form. Equation (\ref{energy_gain_ion_frame_pre_final_s_g0}) with $M=m_{\rm e}$, $Q=q_{\rm e}=-e$ leads immediately to the classic result
\begin{equation}
W^{\rm max}\big|_{g=0} = m_{\rm e} (4\gamma^3 - 3\gamma - 1)
\end{equation}
for the maximum energy gain of a test electron in a plasma wakefield accelerator~\cite{esarey:1995, esarey:2009}. Using (\ref{s_def}) and (\ref{l_approx}) in the Appendix, it follows that $\pi s/2 \approx \sigma$ where
\begin{equation}
\label{sigma_def_no_c}
\sigma = \frac{\sqrt{2}\,m_\alpha \gamma^{3/2}}{\omega_p}
\end{equation}
and hence
\begin{equation}
\label{energy_gain_ion_frame_pre_final_s_no_c}
W^{\rm max} \approx W^{\rm max}\big|_{g=0} \bigg[1 - \frac{g^2 B^2}{m_\alpha^2}\bigg(1-\frac{\tanh\sigma}{\sigma}\bigg)\bigg].
\end{equation}
The starting point in Section~\ref{sec:results} is obtained by restoring appropriate powers of $c$, $\mu_0$, $\hbar$ in (\ref{energy_gain_ion_frame_pre_final_s_g0}), (\ref{sigma_def_no_c}), (\ref{energy_gain_ion_frame_pre_final_s_no_c}).
%%%
\section{Conclusion}
%%%%
Axions are hypothetical particles that solve the strong CP problem and are one of the most promising candidates for Cold Dark Matter; it is conceivable that they pervade the cosmos. In addition to QCD axions, type IIB string theory predicts a plethora of axion-like particles (ALPs) and we have unearthed some of the implications of ALPs for plasma-based wakefield acceleration in ultra-strong magnetic fields. Typical values of the ALP mass and ALP-photon coupling strength suggest that minimally-coupled ALPs generated within the plasma wave could dramatically suppress the effectiveness of the acceleration process in ultra-strong field environments. While such suppression is unlikely to play a significant role in scenarios that exploit moderate magnetic fields, such as those in references~\cite{chen:2002, ebisuzaki:2014}, the theoretical significance of axions, and ALPs in general, suggests that such particles should be taken into account in any attempt to explain ultra-high energy cosmic rays using plasma-based wakefield acceleration in the strongest magnetic fields in the Universe.

Our analysis does not include the evolution of the the driver at the head of the wakefield, ALPs directly generated by the electromagnetic field of the driver, or the ALPs generated by external sources; such developments could form the basis of a future study.
%%%%
\section*{Acknowledgements}
%%%%
This work was undertaken as part of the ALPHA-X consortium funded under EPSRC grant EP/J018171/1. AN is supported by the ELI-NP Project and the European Commission FP7 projects Laserlab-Europe (Grant 284464) and EuCARD-2 (Grant 312453). DAB and AN are also supported by the Cockcroft Institute of Accelerator Science and Technology  (STFC grant ST/G008248/1). There are no additional data files for this article.
%%%%%

%%%
\section*{Appendix: The value of $\mu_*$}
%%%%
The maximum amplitude electrostatic wave in the absence of quantum vacuum polarisation and the ALP field is governed by
\begin{equation}
\label{nu_prime}
\frac{d\nu}{d\zeta}= \sqrt{2} \gamma^2 \omega_p (v\sqrt{\nu^2-\gamma^2} - \nu + \gamma)^{1/2}
\end{equation}
where the electric field is given by (\ref{E_mu}) with $\mu$ replaced by $\nu$.
It is straightforward to obtain an approximate analytical solution to (\ref{nu_prime}) and determine a good approximation to the Fourier coefficients $\alpha_n$ of the ALP field when $\gamma \gg 1$. Inspection of (\ref{nu_prime}) reveals that the maximum and minimum values of $\nu$ are $\gamma^3(1+v^2)$, $\gamma$ respectively and, in the following, we will choose the constant of integration in (\ref{nu_prime}) such that $\nu = \gamma$ at $\zeta=0$.

The phase $\zeta$ of the wave can be written as
\begin{align}
\nonumber
\zeta(\nu) &= \frac{1}{\sqrt{2} \gamma^2 \omega_p} \int^\nu_\gamma \frac{1}{(v\sqrt{\chi^2 - \gamma^2} - \chi + \gamma)^{1/2}}\,d\chi\\
\label{zeta_integral}
&= \frac{1}{\sqrt{2\gamma} \omega_p} \int^{\bar{\nu}}_{\gamma^{-2}} \bigg[\sqrt{\bigg(1-\frac{1}{\gamma^2}\bigg)\bigg(\bar{\chi}^2 - \frac{1}{\gamma^4}\bigg)} - \bar{\chi} + \frac{1}{\gamma^2}\bigg]^{-1/2}\,d\bar{\chi}
\end{align} 
where $\bar{\nu} = \nu/\gamma^3$, $\bar{\chi} = \chi/\gamma^3$ and $v=\sqrt{1-\gamma^{-2}}$ has been used to clearly show the $\gamma$ dependence of the integrand.
However,
\begin{equation}
\sqrt{\bigg(1-\frac{1}{\gamma^2}\bigg)\bigg(\bar{\chi}^2-\frac{1}{\gamma^4}\bigg)} - \bar{\chi} + \frac{1}{\gamma^2} = \frac{1}{\gamma^2}\bigg(1-\frac{\bar{\chi}}{2}\bigg) + {\cal O}(\gamma^{-4})
\end{equation}
and hence
\begin{align}
\nonumber
\zeta &\approx \sqrt{\frac{\gamma}{2}} \frac{1}{\omega_p} \int^{\bar{\nu}}_0 \frac{1}{\sqrt{1-\bar{\chi}/2}}\,d\bar{\chi}\\
\label{zeta_approx}
& = \sqrt{\frac{\gamma}{2}}\frac{4}{\omega_p}\bigg(1-\sqrt{1-\frac{\bar{\nu}}{2}}\bigg)
\end{align}
is the dominant behaviour of (\ref{zeta_integral}) when $\gamma \gg 1$. The period $l$ of the oscillation is
\begin{equation}
l = 2\zeta\big|^{\nu = \gamma^3(1+v^2)}_{\nu = \gamma}
\end{equation}
and thus
\begin{equation}
\label{l_approx}
l \approx 2\zeta\big|^{\bar{\nu} = 2}_{\bar{\nu}=0} = \frac{4\sqrt{2\gamma}}{\omega_p}
\end{equation}
follows from (\ref{zeta_approx}). Solving (\ref{zeta_approx}) for $\nu$ and writing the result in terms of $l$ using (\ref{l_approx}) yields
\begin{equation}
\label{nu_approx}
\nu(\zeta) \approx 8\gamma^3 \frac{\zeta}{l}\bigg(1-\frac{\zeta}{l}\bigg) \quad \mbox{for $0 \le \zeta \le l$}
\end{equation}
when $\gamma \gg 1$. 

Thus, the coefficients $\nu_n$ of the Fourier series
\begin{equation}
\nu(\zeta) = \sum\limits^{\infty}_{n=-\infty} \nu_n \exp\bigg(2\pi i n \frac{\zeta}{l}\bigg)
\end{equation}
follow immediately:
\begin{align}
\label{nu_n}
\nu_n &= \frac{1}{l} \int^l_0 \exp(-2\pi i n \zeta/l)\,\nu(\zeta)\,d\zeta\\
\label{nu_n_approx}
&\approx
\begin{cases}
\displaystyle\frac{4\gamma^3}{3}\quad&\mbox{for $n=0$}\\
\displaystyle-\frac{4\gamma^3}{\pi^2 n^2}\quad&\mbox{for $n\neq 0$}
\end{cases}
\end{align}
when $\gamma \gg 1$.

Although (\ref{nu_approx}) is a poor approximation to $\nu$ very close to the minima of $\nu$, numerical investigation reveals that the difference between the approximate value of (\ref{nu_n}) and the exact value is less than $1\%$ for $|n| \in \{0,1,2,3\}$ with $\gamma = 10$. The percentage error slowly increases with $|n|$, but the Fourier coefficient $\nu_n$ rapidly decays with $n$ and the absolute error is negligible. Moreover, the accuracy of the approximation increases with $\gamma$; laser-driven plasma wakefields have a Lorentz factor $\gamma$ in the range $10 - 100$ and the Lorentz factor of laboratory-based electron-driven plasma wakefields are even higher ($\gamma \sim 10^5$). For notational convenience we will henceforth treat (\ref{nu_n_approx}) as an equality rather than approximate equality. 

We now turn to the value of the remnant $\varphi[\nu]$ of the ALP field in (\ref{integrated_tauK_e2_ODE_axion_eliminated_approx}). Inspection of (\ref{nu_n_approx}) and the expression
\begin{equation}
\label{alpha_n_nu}
\alpha_n = \frac{gB l}{4\pi^2 n^2 + m^2_\alpha \gamma^2 l^2}\frac{m_{\rm e}}{q_{\rm e}} 2\pi i n \nu_n
\end{equation}  
obtained from (\ref{alpha_n}) reveals that $\alpha_{-n} = -\alpha_n$ and so, using (\ref{alpha_fourier}),
\begin{align}
\label{alpha_0}
&\alpha(0) = \sum_{n=-\infty}^{\infty} \alpha_n = 0,\\
\label{alpha_l2}
&\alpha(l/2) = \sum_{n=-\infty}^{\infty} \alpha_n (-1)^n = 0.
\end{align}
It is also straightforward to show that
\begin{align}
\label{alpha_prime_0}
&\alpha^\prime(0) = \sum_{n=-\infty}^{\infty} \alpha_n \frac{2\pi i n}{l} = \kappa_e + \kappa_o,\\
\label{alpha_prime_l2}
&\alpha^\prime(l/2) = \sum_{n=-\infty}^{\infty} \alpha_n \frac{2\pi i n}{l} (-1)^n =\kappa_e - \kappa_o
\end{align}
where $\kappa_e$, $\kappa_o$ result from splitting the Fourier sums into even and odd indices, respectively:
\begin{align}
\label{kappa_e}
&\kappa_e = gB \frac{m_{\rm e}}{q_{\rm e}} \frac{4\gamma^3}{\pi^2} \sum\limits_{\text{$n$ even, $n\neq 0$}}\frac{1}{n^2 + s^2},\\
\label{kappa_o}
&\kappa_o = gB \frac{m_{\rm e}}{q_{\rm e}} \frac{4\gamma^3}{\pi^2} \sum\limits_{\text{$n$ odd}} \frac{1}{n^2 + s^2}.
\end{align}
The details of (\ref{kappa_e}), (\ref{kappa_o}) follow from (\ref{nu_n_approx}), (\ref{alpha_n_nu}) with the parameter $s$ given as
\begin{equation}
s = \frac{m_\alpha \gamma l}{2\pi}.
\end{equation}
Furthermore, the summations in (\ref{kappa_e}), (\ref{kappa_o}) can be expressed in closed form:
\begin{align}
\label{n_even_sum}
&\sum\limits_{\text{$n$ even, $n\neq 0$}}\frac{1}{n^2 + s^2} =
\frac{\pi\coth(\pi s/2) s - 2}{2 s^2},\\
\label{n_odd_sum}
&\sum\limits_{\text{$n$ odd}} \frac{1}{n^2 + s^2} = \frac{\pi\tanh(\pi s/2)}{2 s}.
\end{align}

Evaluating (\ref{axion_remnant_functional}) at $\mu=\nu$ and using (\ref{alpha_0}), (\ref{alpha_l2}), (\ref{alpha_prime_0}), (\ref{alpha_prime_l2}) leads to
\begin{equation}
\label{fnu}
\varphi[\nu]\big|^{\zeta_{\rm II}}_{\zeta_{\rm I}} =
\begin{cases}
2\kappa_e \kappa_o/\gamma^2\,\,&\mbox{if $Q/q_{\rm e} > 0$}\\
- 2\kappa_e \kappa_o/\gamma^2\,\,&\mbox{if $Q/q_{\rm e} < 0$}
\end{cases}
\end{equation}
where
\begin{equation}
\zeta_{\rm I} =
\begin{cases}
0\,\,&\mbox{if $Q/q_{\rm e} > 0$}\\
l/2\,\,&\mbox{if $Q/q_{\rm e} < 0$}
\end{cases}
\end{equation}
and
\begin{equation}
\zeta_{\rm II} =
\begin{cases}
l/2\,\,&\mbox{if $Q/q_{\rm e} > 0$}\\
l\,\,&\mbox{if $Q/q_{\rm e} < 0$}
\end{cases}
\end{equation}
have been used, and the periodicity of the fields has been exploited.

Finally, noting that $\kappa_e \kappa_o = {\cal O}(g^2)$, equation (\ref{mu_II_explicit}) follows from a perturbative analysis of (\ref{integrated_tauK_e2_ODE_axion_eliminated_approx}). Equations (\ref{integrated_tauK_e2_ODE_axion_eliminated_approx}), (\ref{fnu}) yield
\begin{equation}
\label{mu_II_approx}
\mu_* \approx \gamma^3(1+v^2) - \frac{4}{\omega_p^2} \frac{q^2_{\rm e}}{m^2_{\rm e}} \kappa_e \kappa_o
\end{equation}
to lowest order in the ALP-photon coupling constant $g$, and substituting (\ref{kappa_e}), (\ref{kappa_o}), (\ref{n_even_sum}), (\ref{n_odd_sum}) in (\ref{mu_II_approx}) leads to (\ref{mu_II_explicit}).

Although only the leading order dependence on $\gamma$ has been retained in the ${\cal O}(g^2)$ term in (\ref{mu_II_approx}), we have retained the exact $\gamma$-dependence of the remaining terms for accuracy.  
\end{document}